\documentclass[twocolumn,tighten,trackchanges,twocolappendix]{aastex631}

\newcommand{\Msun}{\mathrm{M}_{\odot}}

\newcommand{\numsample}{491}

\usepackage{color}

\shortauthors{Hamilton-Campos et al.}

\graphicspath{{./}{figures/}}

\begin{document}

\title{The Physical Thickness of Stellar Disks to z $\sim$ 2}

\correspondingauthor{Kathleen Hamilton-Campos}
\email{khamil42@jhu.edu}

\author[0000-0003-1296-8775]{Kathleen A. Hamilton-Campos}
\affiliation{William H. Miller III Department of Physics and Astronomy \\ 
Johns Hopkins University \\
3400 N. Charles Street\\
Baltimore, MD 21218, USA \\}
\affiliation{Space Telescope Science Institute \\
3700 San Martin Drive \\
Baltimore, MD 21218, USA \\}

\author[0000-0002-6386-7299]{Raymond C. Simons}
\affiliation{Space Telescope Science Institute \\
3700 San Martin Drive \\
Baltimore, MD 21218, USA \\}
\affiliation{University of Connecticut \\
196 Auditorium Road \\
Storrs, CT 06269, USA \\}

\author[0000-0003-1455-8788]{Molly S. Peeples}
\affiliation{William H. Miller III Department of Physics and Astronomy \\ 
Johns Hopkins University \\
3400 N. Charles Street\\
Baltimore, MD 21218, USA \\}
\affiliation{Space Telescope Science Institute \\
3700 San Martin Drive \\
Baltimore, MD 21218, USA \\}

\author[0000-0001-7472-3824]{Gregory F. Snyder}
\affiliation{Space Telescope Science Institute \\
3700 San Martin Drive \\
Baltimore, MD 21218, USA \\}

\author[0000-0001-6670-6370]{Timothy M. Heckman}
\affiliation{William H. Miller III Department of Physics and Astronomy \\ 
Johns Hopkins University \\
3400 N. Charles Street\\
Baltimore, MD 21218, USA \\}

\begin{abstract}

In local disk galaxies such as our Milky Way, older stars generally inhabit a thicker disk than their younger counterparts. Two competing models have attempted to explain this result: one in which stars first form in thin disks that gradually thicken with time through dynamical heating, and one in which stars form in thick disks at early times and in progressively thinner disks at later times. We use a direct measure of the thicknesses of stellar disks at high redshift to discriminate between these scenarios. Using legacy \emph{HST} imaging from the CANDELS and GOODS surveys, we measure the rest-optical scale heights of \numsample{} edge-on disk galaxies spanning $0.4 \leq z \leq 2.5$. We measure a median intrinsic scale height for the full sample of 0.74 $\pm$ 0.03 kpc, with little redshift evolution of both the population median and scatter. The median is consistent with the thick disk of the Milky Way today ($0.6\,-\,1.1$ kpc), but is smaller than the median scale height of local disks ($\sim$1.5 kpc) which are matched to our high-redshift sample by descendant mass. These findings indicate that (1) while disks as thick as the Milky Way's thick disk were in place at early times, (2) to explain the full disk galaxy population today, the stellar disks in galaxies need to on average physically thicken after formation.
\end{abstract}


\section{Introduction} \label{sec:intro}

In present-day disk galaxies like our own Milky Way, older stars are generally found at larger distances above and below the galaxy midplane than their younger counterparts \citep{1995AJ....110.2771W, 2006AJ....131..226Y,  2009A&A...501..941H, leaman17}. They are also ``kinematically hotter'' with larger vertical velocity dispersions---on average, the orbit of an older star will loft further from the midplane than that of a younger star. These two observational facts are closely related. 

The collection of older stars in the Milky Way comprises what is known as its ``thick disk'' (with a scale height of $\sim$ 1 kpc), while its gas and younger stars comprise its flatter ``thin disk'' (scale height $\sim$ 270 pc) (\citealt{1983MNRAS.202.1025G, 2016ARA&A..54..529B} and references therein). In reality, the relation between disk thickness (or vertical velocity dispersion) and stellar age in the Milky Way is likely a continuum \citep{bovy12}---with younger stars comprising thinner, kinematically-colder components of the disk and older stars comprising thicker, kinematically-hotter components of the disk. Most disk galaxies in the local universe appear to have a thick(-er) disk component made up of old(-er) stars \citep{2006AJ....131..226Y}. Understanding when and how the older, thicker components of today's stellar disks developed is an open question---and a key piece of our story of disk galaxy formation. 

The classic explanation for thick disks starts by assuming that the stars that comprise today's thick disks {\emph{first formed in thin disks}}. Those initially-thin disks are then thought to have thickened with time through the dynamical heating of the vertical components of the stellar orbits. That heating could come from gradual internal processes, such as vertical ``kicks'' from gravitational interactions between the stars and asymmetries in the disk (e.g., spiral arms and Giant Molecular Clouds; \citealt{1985ApJ...290...75V}). The heating could also come from fast external processes, such as galaxy-galaxy mergers \citep{wyse06}. It is perhaps reasonable to expect that the older stars---having spent a longer time in the disk with more opportunities for such encounters---would comprise a kinematically hotter and physically thicker component of the disk.

A more recent explanation (one that contrasts the historical picture above) contends that today's thick stellar disks {\emph{were first formed thick}}. This idea is motivated by observations that the velocity dispersion of the ionized gas in high-redshift galaxies (back to $z\,\sim\,3$ or 11.5 Gyr in lookback time) was up to a few times higher than it is in today's galaxies (e.g., \citealt{2006ApJ...653.1027W, 2007ApJ...660L..35K, 2012ApJ...758..106K, wisnioski15, simons16, 2017ApJ...843...46S, ubler19}). If the velocity dispersion of the ionized gas in these galaxies reflects that of their cold molecular star-forming gas {\emph{and}} the stars that are newly forming inherit the kinematics of the gas from which they form, then we might expect for the young stellar disks of these high redshift galaxies to be physically thicker than the young stellar disks today. In this picture, today's old thick stellar disks are simply the descendants of the young thick disks formed at high redshift. As the velocity dispersion of the ionized gas in galaxies gradually declines with time (as is observed; \citealt{2012ApJ...758..106K, 2017ApJ...843...46S, ubler19}), we might expect for later generations of stars to form in progressively thinner disks. 

In brief, the question is whether the {\emph{stars}} that comprise today's disk galaxies: 
\begin{enumerate}
\item formed in thin disks (i.e., near the galaxy midplane) at all cosmic times which subsequently thicken into thick disks, or

\item formed in thick disks (i.e., at large distances above and below the midplane) at early times and in progressively thinner disks at later times.
\end{enumerate}
Both of these scenarios qualitatively reproduce the observed trend in the Milky Way---that older stars comprise progressively thicker portions of the disk.

Numerical simulations of galaxy formation in the context of a $\Lambda$CDM cosmology have made considerable progress in addressing this question (e.g., \citealt{2021MNRAS.503.1815B, 2021MNRAS.502.1433M}), but there is not yet theoretical consensus. For instance, \cite{2021MNRAS.503.1815B} argue that both scenarios are at play---older stars are formed in thicker disks at higher redshifts and are {\emph{also}} then kinematically-heated after birth into even thicker disks. \cite{2021MNRAS.502.1433M} report that the stars in their simulations form in thin disks at all times, but that the continuous rearrangement of the orientation of the star-forming disk leads to a gradual thickening of the disk.

To distinguish between these scenarios, one needs direct measurements of the thicknesses of galaxy disks back to the early universe. At early times, the scenarios offer opposing predictions. The first scenario predicts that disks (as a composite including both the thick and thin components, which is what is observable) should be physically thinner at higher redshift and progressively thicker closer to the present day. The second scenario predicts that the composite disks should be as thick at high redshift as the thick disks today and progressively thinner closer to the present day. 

In this paper, we measure the vertical scale heights of \numsample{} galaxies from $z\sim0.4$ to $z\sim2.5$ to reveal how and when galaxies formed their thick stellar disks. To do that, we use archival \emph{Hubble Space Telescope} (\emph{HST}) IR imaging from the CANDELS and GOODS surveys \citep{2004ApJ...600L..93G, 2011ApJS..197...35G, 2011ApJS..197...36K}. Given the limited resolution of the {\emph{HST}} imaging, we are only able to infer the composite thickness of these external galaxies---we can not distinguish between their thin and thick components. The galaxies in this sample span a stellar mass range of $9 \leq \log M_*/\Msun \leq 11$ (with the majority at $\log M_*/\Msun\leq10$) and a redshift range of $0.4 \leq z \leq 2.5$ (with the majority at $z\,>\,1$). This work builds on previous formative studies using \emph{HST} imaging to measure the disk scale heights of high redshift galaxies in the Ultra Deep Field and Frontier Fields \citep{2006ApJ...650..644E, 2017ApJ...847...14E}. The study presented in this paper includes a number of advances: (1) we study a large sample of galaxies (N = \numsample{}) that allows us to infer the statistics of galaxy populations in redshift,  (2) we focus on a fixed rest-optical wavelengths (0.46--0.66 $\micron$) to mitigate degeneracies between scale height, stellar age, and redshift, and (3) we compare with a population-matched sample of disk galaxies at $z\,=\,0$ to infer evolution.

\begin{figure*}[htb]
     \centering
     \includegraphics[width=\textwidth]{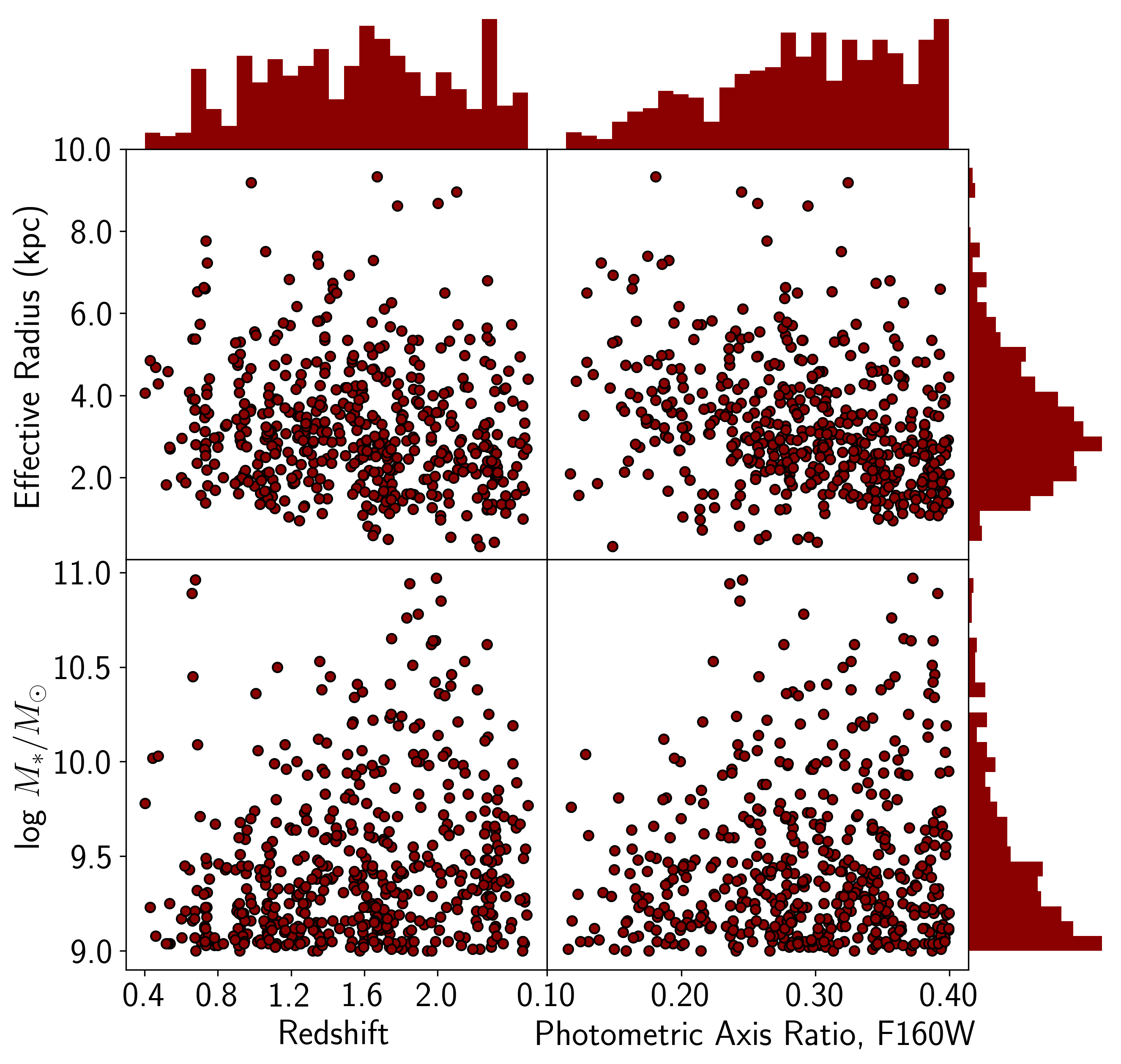}
     \caption{The distributions of stellar mass, redshift, WFC3/F160W effective radius, and WFC3/F160W photometric axis ratio for the galaxies in our sample are shown. Our sample spans a relatively uniform distribution in redshift. The majority of the galaxy sample have low stellar mass ($9\,<\,\log\,M_*/M_{\odot}\,<\,10$). To include galaxies that are sufficiently edge-on, we select galaxies with a WFC3/F160W photometric axis ratio less than $0.4$.}
        \label{fig:hist}
\end{figure*}

In \S \ref{sec:data}, we discuss the {\emph{HST}} imaging used and the selection of our galaxy sample. In \S \ref{sec:methods}, we discuss the measurements of galaxy scale heights from the images. In \S \ref{sec:resdis}, we present our results on the distribution and redshift evolution of scale height. We compare our results against a local sample of disk galaxies that are selected to reflect the expected descendants of our sample in terms of mass, and discuss our findings. In \S \ref{sec:conclude}, we briefly conclude. In Appendix \ref{sec:biases}, we discuss a correction that we apply to the scale heights to account for galaxy inclination. In Appendix \ref{sec:kernel}, we discuss the 1D PSF kernel we use in our model. Where relevant, we adopt a Planck15 $\Lambda$CDM cosmology \citep{2016A&A...594A..13P} with ($h$, $\Omega_m$, $\Omega_\lambda$) = (0.67, 0.31, 0.69).

\begin{figure*}[t]
    \centering
    \includegraphics[width=\textwidth]{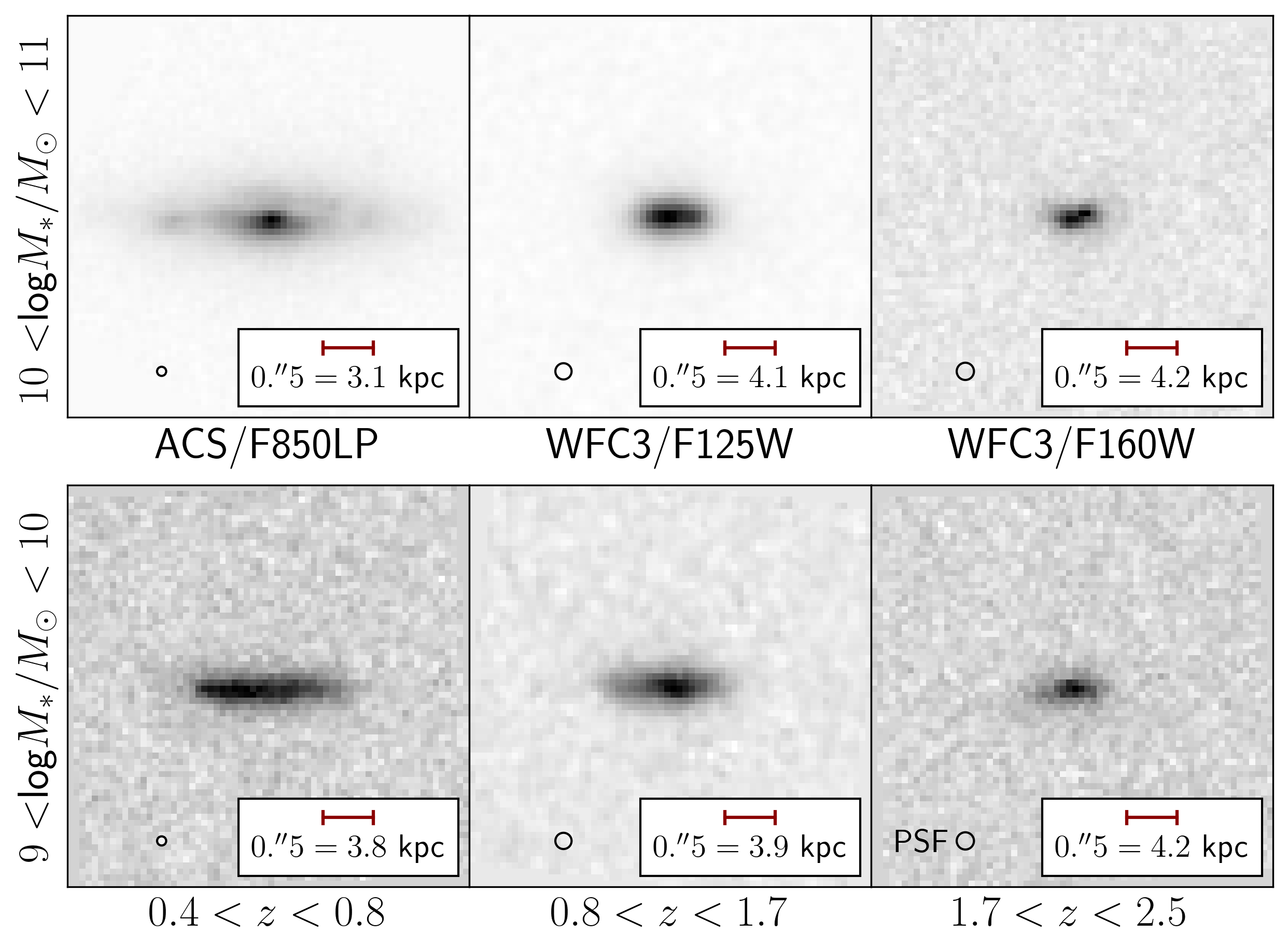}
    \caption{{\emph{HST}}/ACS+WFC3 postage stamps for a random subset of the galaxies in the sample are shown. We select edge-on galaxies using a cut on axis ratio. The top row are high-mass galaxies ($10\,<\,\log\,M_*/M_{\odot}\,<\,11)$ and the bottom row are low-mass galaxies ($9\,<\,\log\,M_*/M_{\odot}\,<\,10)$. Redshift increases from left to right. The filters shown (and used to make the measurements in this paper) vary with redshift to target a fixed rest frame wavelength of $0.46-0.66 \mu m$ (see Figure \ref{fig:rest} and \S \ref{subsec:Select}).  We use ACS/F850LP imaging for galaxies over $0.4 \leq z \leq 0.8$, WFC3/F125W for $0.8 < z \leq 1.7$, and WFC3/F160W for $1.7 < z \leq 2.5$. The postage stamps are $4.\arcsec0$ on a side. A red bar is included to indicate the physical scale. Black circles in the lower left corners represent the point-spread functions for each filter: the full-width half-maximum is the radius of each circle.}
\label{fig:poststamps}
\end{figure*}

\section{Data and Sample Selection} \label{sec:data}
We use archival \emph{Hubble Space Telescope} (\emph{HST}) near-infrared imaging to measure the photometric scale heights of \numsample{} edge-on disk galaxies over $0.4 \leq z \leq 2.5$ in the GOODS-S galaxy field. Here, we discuss the \emph{HST} imaging used, the catalogs containing the physical properties of the galaxies in GOODS-S (\S \ref{subsec:HSTCat}), and the selection of the galaxy sample studied in this paper (\S \ref{subsec:Select}). Figure \ref{fig:hist} shows the distribution of the galaxy sample in mass, size, photometric axis ratio, and redshift. Figure \ref{fig:poststamps} shows \emph{HST} imaging for a random subset of the galaxies in the sample.


\subsection{HST Imaging and Catalogs} \label{subsec:HSTCat}
We use 3-band optical and near infrared imaging of the GOODS-S galaxy field observed with the Advanced Camera for Surveys (ACS/850LP) and the Wide Field Camera 3 (WFC3/F125W+F160W) on \emph{HST}. The imaging was taken as a part of two \emph{HST} Treasury programs: the Cosmic Assembly Near-infrared Deep Extragalactic Legacy Survey (CANDELS; \citealt{2011ApJS..197...35G, 2011ApJS..197...36K}) and the Great Observatories Origins Deep Survey (GOODS; \citealt{2004ApJ...600L..93G}). We use the image mosaics, source detection, segmentation maps, and photometric catalogs of this imaging field as provided by the 3D-HST survey \citep{2014ApJS..214...24S}\footnote{https://archive.stsci.edu/prepds/3d-hst/}. The mosaics include an estimate of the 2D point-spread function, which was constructed by stacking isolated stars in the mosaic \citep{2014ApJS..214...24S}. The catalogs also include estimates of stellar masses as derived from the \texttt{FAST} stellar population synthesis fitting routine \citep{2009ApJ...700..221K}. The \texttt{FAST} fits assume a \citet{2003PASP..115..763C} IMF, \citet{2003MNRAS.344.1000B} stellar population synthesis models, and a \citet{2000ApJ...533..682C}  extinction law. The general uncertainties on stellar masses derived from the  {\emph{HST}} imaging available in GOODS-S are approximately 0.3 dex \citep{2015ApJ...808..101M}. We take the photometric redshifts provided in the 3D-HST catalogs as measured by the \texttt{EAZY} photometric redshift code \citep{2008ApJ...686.1503B}. The typical uncertainty on the photometric redshifts are $0.1\times(1+z)$ (R. C. Simons et al.\ in prep).

We adopt measurements of the F160W photometric axis ratios, position angles, and effective radii of the galaxies in GOODS-S from the \texttt{GALFIT} catalogs described in \cite{2012ApJS..203...24V}. The effective radius is defined as the semi-major axis of the ellipse that contains half of the total light in the best fitting \texttt{GALFIT} single-S\'ersic model. The axis ratio is defined as the ratio of the semi-major and semi-minor axes from the best-fitting S\'ersic model \citep{2012ApJS..203...24V}. 

\subsection{Galaxy Sample Selection} \label{subsec:Select}
To select the galaxies studied in this paper, we use the masses, effective radii, photometric position angles, and photometric redshifts tabulated in the 3D-HST and \texttt{GALFIT} catalogs, as described above.

\begin{figure}
\begin{center}
\includegraphics[width=\columnwidth]{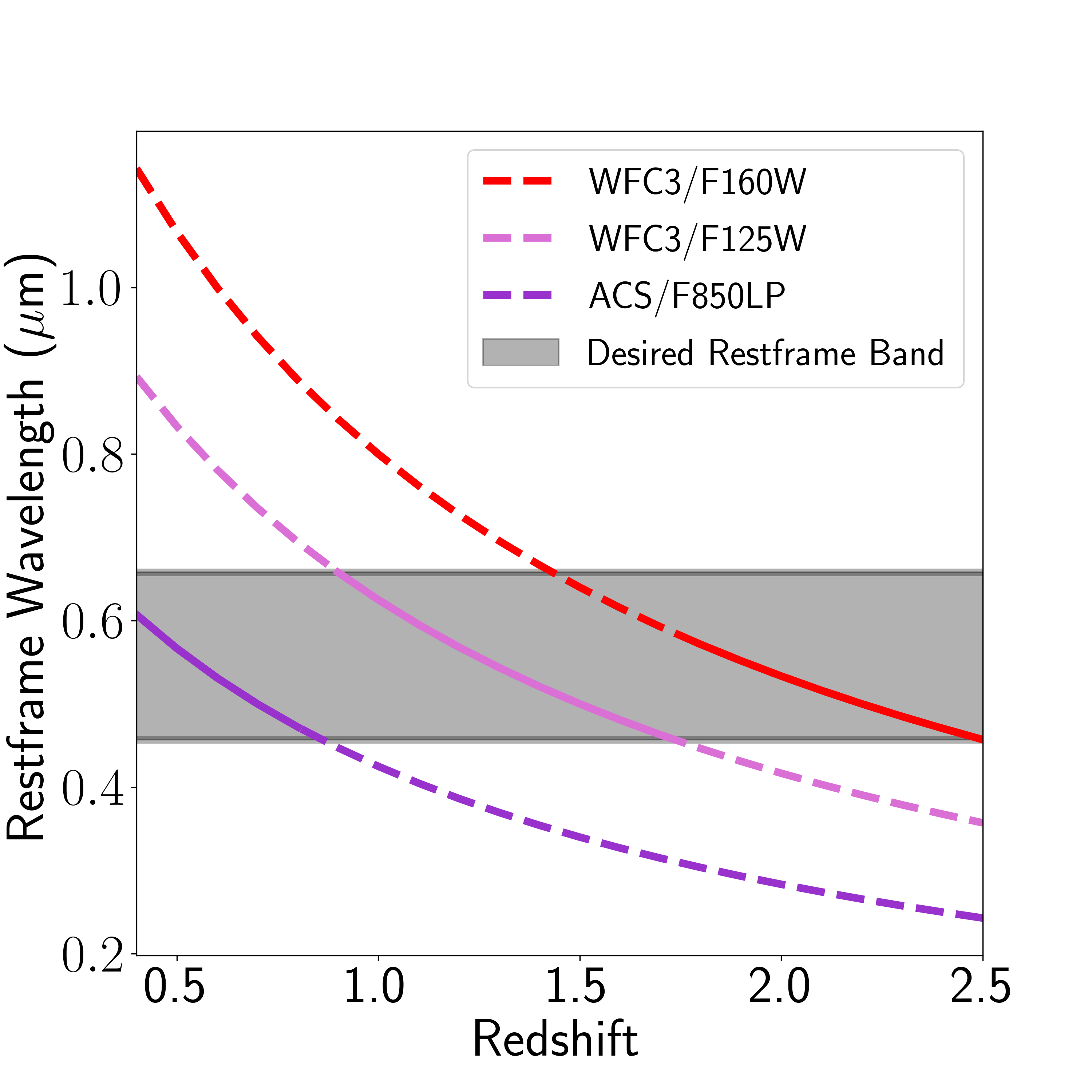}
\caption{We measure the scale heights of the galaxies in the same rest frame wavelength window. This figure shows the redshift evolution of the rest wavelength traced by the characteristic wavelength of the three \emph{HST} imaging filters we use---WFC3/F160W, WFC3/F125W, and ACS/F850LP. The gray shaded region shows our targeted rest wavelength window, spanning 0.46-0.66 $\mu $m. For each galaxy, we carry out measurements in the filter appropriate for its redshift: F160W for $1.7 < z \leq 2.5$, F125W for $0.8 < z \leq 1.7$, and F850LP for $0.4 \leq z \leq 0.8$.}
\label{fig:rest}
\end{center}
\end{figure}

We select a parent sample of 6933 galaxies in GOODS-S that span a fixed range in mass ($\log\,M_*/\mathrm{M}_{\odot}\,=\,9-11$) and photometric redshift ($z_{\mathrm{phot}}\,=\,0.4\,-\,2.5$). The redshift minima and maxima are chosen to bound a fixed rest wavelength window for the {\emph{HST}}/WFC3+filters used in this paper (Figure \ref{fig:rest}). From the parent sample, we use two criteria to select galaxies with a well-defined orientation, i.e., a well-defined major and minor axis, so that we can reliably orient the plane of the galaxy. To do that,  we first keep galaxies that have been flagged with the ``good fit" designation in the \texttt{GALFIT} catalogs. We then compare the position angle measured by the 3D-HST team using the SExtractor code \citep{1996A&AS..117..393B} to that measured from \texttt{GALFIT} for each galaxy. If the two position angles disagree by more than 10$^{\circ}$, we consider the position angle to be uncertain and the galaxy is discarded. Finally, we remove 3 galaxies with errant effective radii ($>5$ arcsec). Together, these criteria reduce the sample to 3032 galaxies.


To mitigate projection effects in the scale height measurements, we ideally want to use galaxies whose disks are oriented perfectly edge-on relative to our line of sight (LOS). In reality, all galaxies possess a finite LOS inclination and so our practical goal is to select galaxies that are {\emph{nearly}} edge-on---and to model and correct for the measurement biases introduced by the mean residual inclination of the sample. We consider a galaxy sufficiently edge-on if it has an F160W photometric axis ratio of $b/a \leq 0.4$. This corresponds to an inclination of $<\,20^{\circ}$ from edge-on for disks with intrinsic axial ratios (commonly denoted $q$) of 0.25. We then use simulations of randomly-inclined galaxy disks to calculate the bias to the median of the measured scale heights due to the residual inclination (see Appendix \ref{sec:biases}). Following the above selection of {\emph{nearly}} edge-on galaxies, our sample is reduced to 1248 galaxies. Finally, we remove galaxies with a sufficiently close neighbor ($\,<\,2\arcsec$) in the full 3D-HST source catalog. The rejected galaxies include those in late-stage encounters and those with chance projections---both of which would challenge our fitting routine.  An additional 36 galaxies were discarded due to bad surface brightness fits (see later, \S \ref{sec:methods}). 

The final sample includes \numsample{} galaxies. The distribution of the sample in size, mass, photometric redshift, and photometric axis ratio is shown in Figure \ref{fig:hist}. The {\emph{HST}} imaging for a random subset of galaxies in the sample is shown in Figure \ref{fig:poststamps}.

\subsection{SDSS Comparison Sample}\label{sec:SDSS}

We compare the measurements of the galaxy scale heights of our sample against those of disk galaxies in the local universe measured from the Sloan Digital Sky Survey (SDSS; \citealt{2003AJ....126.2081A, 2014ApJ...787...24B}). \citet{2014ApJ...787...24B} used SDSS g-band imaging ($\sim$0.41 - 0.53 $\mu$m; characteristic wavelength $\sim$ 0.48 $\mu$m) to measure the scale heights of 4768 local edge-on disks. This wavelength window roughly matches the rest frame window targeted for our sample. For a fair comparison, we want to compare the galaxies in our sample with their anticipated descendants in the local universe. To that end, we downselect the \citet{2014ApJ...787...24B} sample to only include galaxies in the stellar mass range $9.5\,<\,\log\,M_*/M_{\odot}\,<\,10.5$. We calculate the stellar mass of each galaxy in the full \citet{2014ApJ...787...24B} SDSS sample using their absolute g-band magnitudes, redshift-determined distances, and a fixed mass-to-light ratio \citep{1979ARA&A..17..135F}. The down-selected SDSS sample includes 1,679 galaxies and spans a stellar mass range (p$_{\mathrm{16th}}$, p$_{\mathrm{50th}}$, p$_{\mathrm{84th}}$) of $\log\,M_*/M_{\odot}$ = (9.9, 10.2, 10.4). Given galaxy mass growth expectations from abundance matching \citep{moster13}, the galaxies in our sample $\log M_*/\Msun\,\sim\,$ (p$_{\mathrm{16th}}$, p$_{\mathrm{50th}}$, p$_{\mathrm{84th}}$) = (9.1, 9.3, 9.9) are expected to evolve in mass  \citep{2015ApJ...803...26P, 2017ApJ...843...46S} to roughly match the mass range of this down-selected SDSS sample.

\section{Measurements of Galaxy Scale Height} \label{sec:methods}

In this section, we describe the fits to the galaxy scale height from the {\emph{HST}} imaging. 

For each galaxy in the sample, we extract $4\,\times\,4\arcsec$ postage stamps from the ACS/F850LP, WFC3/F125W, and WFC3/F160W 3D-HST imaging mosaics. The stamps are centered on the galaxy center. We extract the corresponding stamps of the inverse variance weight and segmentation maps using their respective mosaics. In each stamp, we use the segmentation map to mask extraneous sources and identify background pixels. We rectify the postage stamps so that the major axis of the galaxy (i.e., the disk midplane, as defined by the F160W photometric position angle) lies along the horizontal direction of the image. Figure \ref{fig:poststamps} shows the rectified stamps for a random subset of 6 galaxies in the final sample. 

We extract vertical surface brightness (and uncertainty) profiles along each column of the postage stamps. The columns are separated by 0.$\arcsec$06---which corresponds to 0.33 kpc at z = 0.4 and 0.52 kpc at z = 2.

We fit each of the observed surface brightness profiles using a model that includes a 1D convolution of the {\emph{HST}} point-spread function (PSF) and a 1D sech$^2$ surface brightness profile,

\begin{equation}
    \mathrm{sech}^{2}(A, \mathrm{z}, \mathrm{z}_{0}) = A \times \frac{4}{(\mathrm{e}^{\Delta \mathrm{z}/\mathrm{z}_{0}}+\mathrm{e}^{-\Delta \mathrm{z}/\mathrm{z}_{0}})^{2}}
    \label{eq:sech2}
\end{equation}

\noindent where $A$, $\Delta \mathrm{z}$, and $\mathrm{z}_{0}$ are the amplitude, the position of the galaxy midplane, and the scale height\footnote{The scale height defined in the sech$^2$ model, and adopted in this paper, is $\sim$one-half of the exponential scale height.}, respectively. We construct and use a 1D convolution kernel appropriate for each imaging filter (see Appendix \ref{sec:kernel}). The sech$^2$ profile assumes that the stars are distributed above and below the disk isothermally \citep{1988A&A...192..117V, 2011ARA&A..49..301V}, with a number density that decreases with the distance from the midplane as $ n(\mathrm{z}) = n_0 \times \mathrm{sech}^{2}(\mathrm{z}/\mathrm{z}_{0})$, where $n_0$ is the number density of stars in the galaxy midplane. We assume that the galaxies have an intrinsic geometry of a disk. A number of studies have argued from both an empirical \citep{Ravindranath06, vdw14, Zhang19} and theoretical perspective \citep{Ceverino15, Tomassetti16, Pandya19}, that the majority of the galaxy population at $z\,>\,1$ and $\log\,M_*/\mathrm{M}_{\odot}\,<\,10$ (which comprises the bulk of our sample, Figure \ref{fig:hist}) have stellar structures that are intrinsically elongated (i.e. prolate) and not disky. For galaxies in our sample that are intrinsically prolate, the ``scale height" that we measure represents the physical thickness of the short axis of the system.

For each galaxy in our sample, we measure the scale height using the imaging band that covers a fixed rest wavelength window ($0.46-0.66\,\mu\mathrm{m}$) at the redshift of the source (Figure \ref{fig:rest}). This allows us to track the same portion of the rest frame SED across our full sample of galaxies. In doing so, we mitigate pseudo-evolution that might arise from a dependence between stellar population age and scale height at fixed redshift. To first order, this fixed portion of the SED is contributed to by similarly-aged stellar populations at different redshifts. The types (i.e., ages and evolutionary states) of stellar populations that contribute light at this rest frame wavelength window will depend somewhat on the star-formation rate history. Typically, this portion of the spectrum ($\sim$0.5\,$\mu$m) will be dominated by $>9$ Gyr old main-sequence stars \citep{conroy13}. We do not consider the effect of dust attenuation on the observed surface brightness profiles. Dust tends to lie in the midplanes of galaxies and will suppress midplane light. For a galaxy with high midplane dust obscuration, the observed vertical surface brightness profile will be broadened and the disk would appear thicker than its intrinsic thickness. Using imaging from the SDSS survey, \citet{2014ApJ...787...24B} inferred that dust attenuation leads to a $<15\%$ increase in the observed disk thickness in local galaxies. Without an empirical constraint on the vertical distribution (and attenuation properties) of the dust in our high-redshift sample, we exclude the minor correction for dust. 

For each column of each galaxy, we employ the Bayesian Markov chain Monte Carlo (MCMC) Python package \texttt{emcee} \citep{2013PASP..125..306F} to derive the probability distribution of the scale height. The model of each profile (Eq. \ref{eq:sech2}) has three free parameters: the amplitude, the position of the galaxy midplane, and the scale height. We adopt a top-hat prior for each. The lower and upper bounds of the amplitude prior are set to 0.5$\times$ and 2.5$\times$ the maximum surface brightness of the profile. The bounds of the midplane prior are set to 3 pixels above or below, respectively, the location of the maximum surface brightness. Finally, the lower and upper bounds of the scale height prior are set to 0 pixels and 10 pixels (0.\arcsec6), respectively. We run \texttt{emcee} using 100 walkers and 2000 steps per walker. We discard the first 100 steps for ``burn-in". A small ($\sim5\%$) fraction of the fits are discarded due to unreliable results---generally surface brightness profiles with unflagged contamination for which the posterior of the scale height is either unconstrained or diverges to unrealistically large values.

\begin{figure}
    \centering
    \includegraphics[width=\columnwidth]{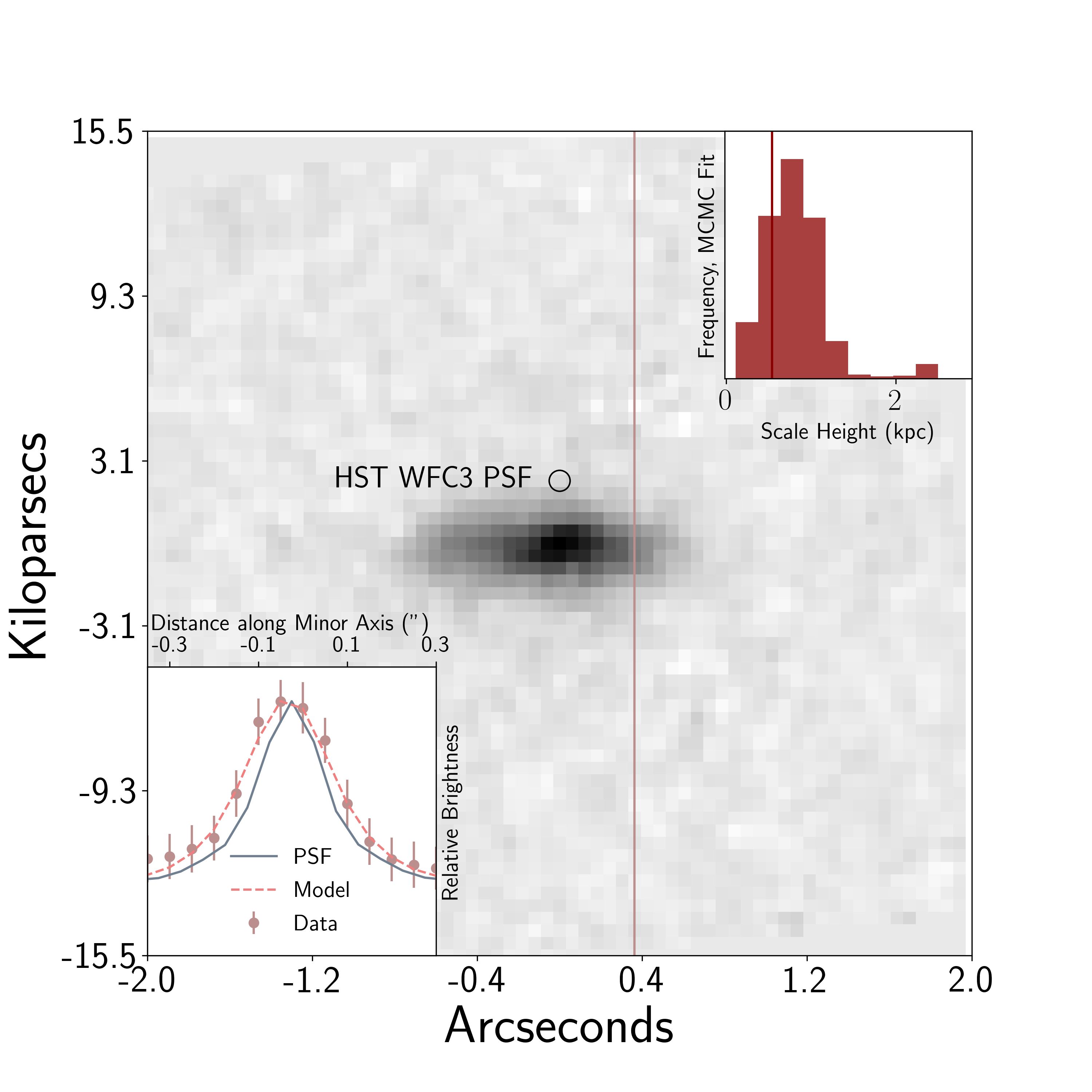}
    \caption{The WFC3/F125W image and fitting results for a single galaxy in the sample is shown. For consistency among galaxies, we measure the scale height as the values at the effective radius. This column is shown as a light brown vertical line. The histogram in the upper right shows the results of our MCMC fitting process for the effective radius column. The final reported scale height, the 50th percentile of these values, is marked by the vertical dark red line. Above the center of the galaxy, we show the \emph{HST} WFC3 PSF as a grey circle whose diameter is the full-width half-maximum of the PSF. We compare this to the data with associated error bars (light brown) and the model (dashed red line) in the lower left corner. This plot demonstrates we measure broadening above and beyond the nominal PSF for our galaxies.}
    \label{fig:diag}
\end{figure}

Figure \ref{fig:diag} shows the results of the fit for a single column of an example galaxy ($\log\,M_*/M_{\odot}\,=\,9.5$ at $z = 0.90$), which rests near the median redshift and stellar mass of the sample. The vertical surface brightness profile shown is measured at one effective radius from the center of the galaxy. The effective radius is marked by the vertical line. We show a circle indicating the full-width at half-maximum of the 2D PSF of the {\emph{HST}}/WFC3 image. In the top right panel, the posterior distribution of the scale height is shown, and the 50$^{\mathrm{th}}$ percentile is marked. The bottom left panel shows the vertical surface brightness profile and its uncertainty, the 50$^{\mathrm{th}}$ percentile model, and the 1D PSF of the image. We note that the observed surface brightness profile is broadened beyond the 1D PSF. This indicates a physical thickness that is detectable with the \emph{HST} image. Repeating this process for each column of each galaxy, we derive the radial profiles of the scale height for all 491 galaxies in the sample.

For simplicity, in what follows we present a single scale height measurement for each galaxy. Specifically, we report the {\emph{weighted-average and uncertainty}} of the two posterior distributions derived at one effective radius from either end of the galaxy center.

\begin{figure*}[ht]
\begin{center}
\includegraphics[width=\textwidth]{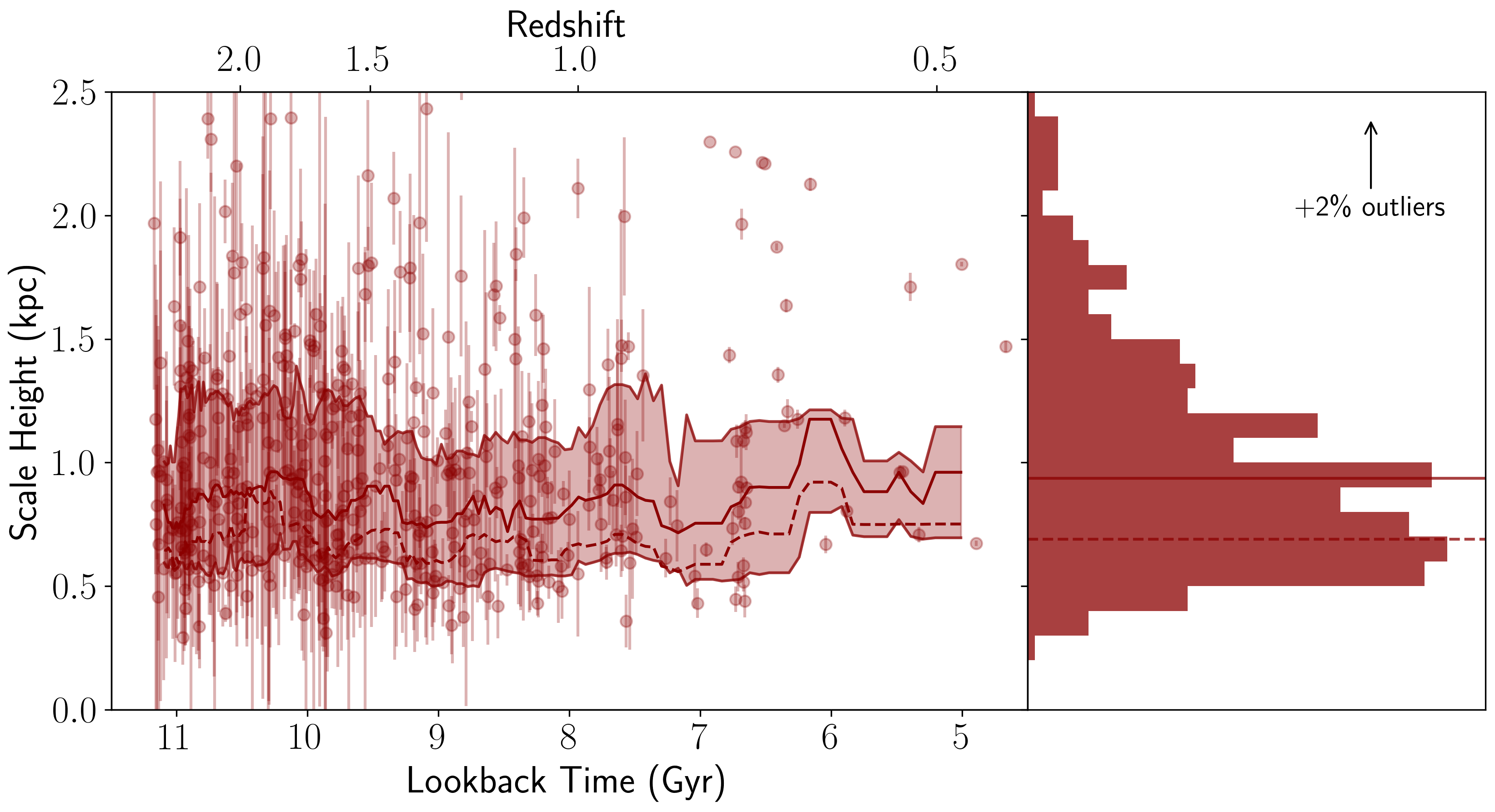}
\caption{The scale heights of the galaxy sample are shown as a function of lookback time (bottom x-axis) and redshift (top x-axis). The scale height is measured at one effective radius from the galaxy center. The weighted average and uncertainty of the two measurements carried out on both sides of the galaxy are reported. The running median (solid red line) and 16$^{\mathrm{th}}$--84$^{\mathrm{th}}$ percentiles (shaded region) of the sample are shown. The dashed red line shows the median corrected for inclination (see Appendix \ref{sec:biases}). The median galaxy scale height of the sample is generally constant over a large period of cosmic time ($0.4 \leq z \leq 2.5$). We truncate the vertical axis at 2.5 kpc, and note that 2\% of the sample have larger scale heights and are not shown here. In the right panel, we show the distribution of the sample, with the median and inclination-corrected median marked.}
\label{fig:scales}
\end{center}
\end{figure*}

\begin{figure*}[htb]
\begin{center}
\includegraphics[width=\textwidth]{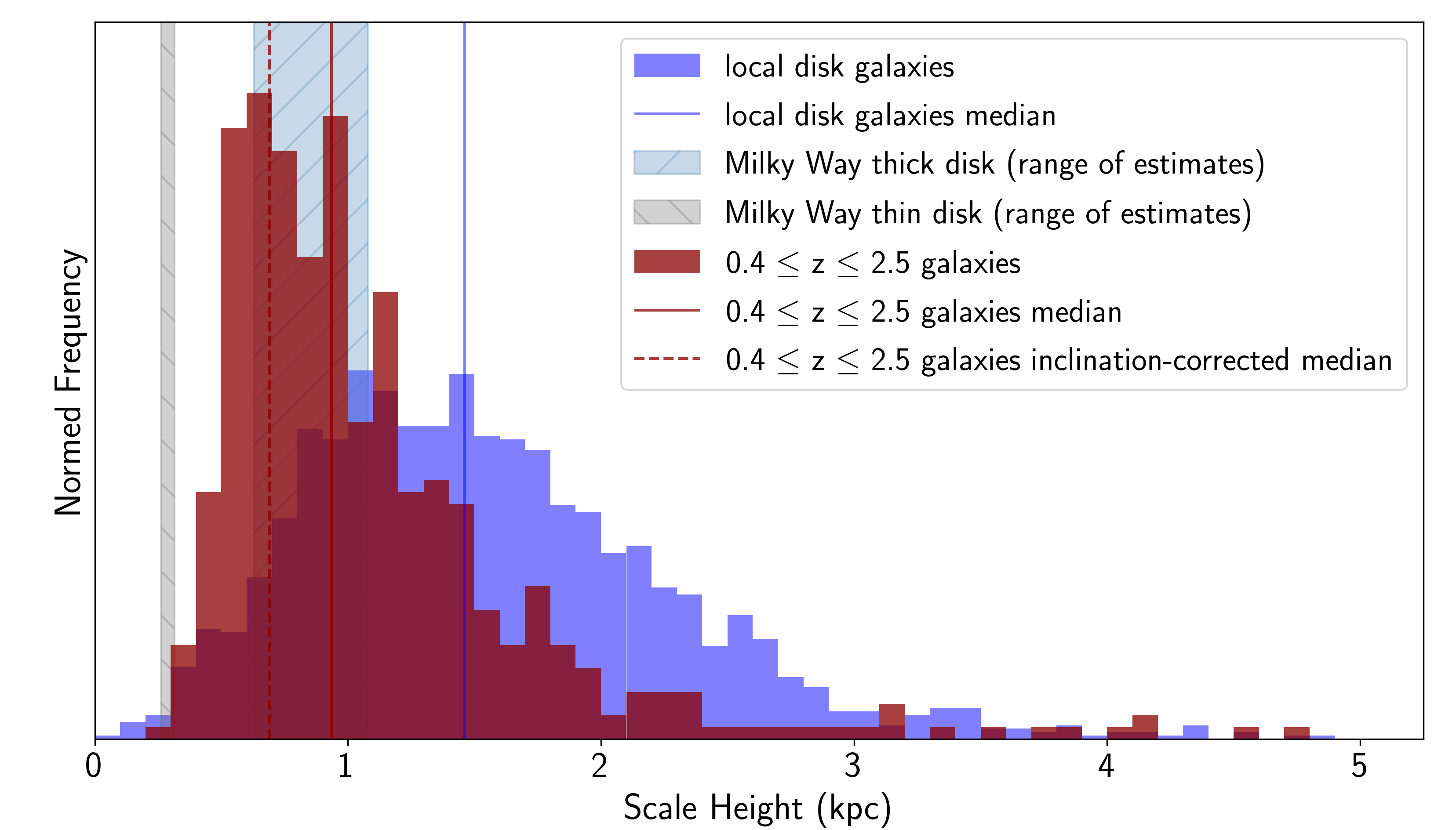}
\caption{The distribution of the galaxy scale heights of our sample is shown in red. The sample spans $0.4\,\leq z\,\leq2.5$. The measurements are carried out at one effective radius from the center of each galaxy. The distribution of a population-matched sample of disk galaxies in the local universe (SDSS; \citealt{2014ApJ...787...24B}) is shown in blue. The median and inclination-corrected median of our high redshift sample are shown with solid and dashed red lines, respectively. The median of the local sample is indicated with a vertical blue line. The range of estimates for the thick (light blue) and thin (grey) disks of the Milky Way are also shown (\citealt{2016ARA&A..54..529B} and references therein). The inclination-corrected median scale height of the high-redshift galaxies ($\sim750$ pc) is smaller than that of the disk galaxies today ($\sim1500$ pc) but is similar to that of the thick disk of the Milky Way ($\sim600\,-1100$ pc).}
\label{fig:today}
\end{center}
\end{figure*}

\section{Results and Discussion} \label{sec:resdis}

In Figures \ref{fig:scales} and \ref{fig:today}, we present the vertical scale heights as measured at one effective radius for \numsample{} galaxies spanning a redshift range of $0.4\,\leq z \leq\,2.5$ and a stellar mass range of $9 \leq \log M_*/\Msun \leq 11$. The majority of the sample (74$\%$) lie at $z>1$ and $\log M_*/\Msun\,<\,10$. As discussed in \S \ref{subsec:Select}, the measurements are carried out at a fixed rest frame wavelength of $0.46\,-\,0.66\,\mu\mathrm{m}$---mitigating potential differences between older (redder) and younger (bluer) stellar populations at a fixed redshift. 

In the left panel of Figure \ref{fig:scales}, we show the scale heights and uncertainties of the full sample as a function of redshift (lookback time). The majority of the sample are consistent with a measurable finite thickness---$\gtrsim97\%$ ($\gtrsim84\%$) of the sample have scale heights that are at least 1$\sigma$ (2$\sigma$) larger than zero. The red-shaded region shows the median and the $16^{\mathrm{th}}-84^{\mathrm{th}}$ percentile span of the sample running with lookback time. The solid red line shows the as-is measured median of the population, and the dashed red line shows the median corrected for the average (small, but non-zero) inclination of the galaxy sample. For the latter, we apply a correction factor of 22\%---see Appendix \ref{sec:biases} for details. In the right panel of Figure \ref{fig:scales}, we show the distribution of scale heights of the full sample with the median and inclination-corrected median marked---0.94 ($\pm$0.04; standard error on the median) kpc and 0.74 ($\pm$0.03) kpc, respectively. 

We do not find evidence for an evolution in either the median or the scatter of the population with redshift---the red shaded region in the left panel of Figure \ref{fig:scales} is flat. For the full sample, we measure a span of 0.6 to 1.4 kpc ($16^{\mathrm{th}}-84^{\mathrm{th}}$ percentiles) or a 1$\sigma$ scatter of 0.35 kpc.

We split the sample into three redshift bins. At high ($1.7 < z \leq 2.5$) and intermediate ($0.8 < z \leq 1.7$) redshifts, the as-is median scale heights of the sample are 0.96 $\pm$ 0.07 kpc and 0.88 $\pm$ 0.04 kpc, respectively. At low redshift ($0.4 \leq z \leq 0.8$)---where the sample is sparse---the median scale height is 1.15 $\pm$ 0.15 kpc. The inclination-corrected median scale heights of the three bins are 0.75 ($\pm\,$0.05),  0.69 ($\pm\,$0.03), and 0.90 ($\pm\,$0.12) kpc, respectively. See Table \ref{table:z0_results} for a summary.

We compare our results with previous studies \citep{2006ApJ...650..644E, 2017ApJ...847...14E} that measured the scale heights of high redshift galaxies with {\emph{HST}}/ACS imaging in the Hubble Ultra Deep Field and Frontier Field Parallels. For context, we note that these measurements were carried out using bluer filters (F435W, F606W, F775W, and F850LP) than those used in this paper and are not corrected for inclination. In total, these samples include $\sim$200 galaxies spanning $7 \leq \log M_*/\Msun \leq 12$ and redshifts $0.5\leq z \leq 4.5$. Over this redshift range, \citet{2006ApJ...650..644E} report an average scale height of 1 $\pm$ 0.4 kpc. Binning by redshift, \citet{2017ApJ...847...14E} report an average of 1.03 $\pm$ 0.25 kpc at $0.5 < z < 1.5$ and 0.63 $\pm$ 0.24 kpc at $1.5 < z < 2.5$. Combining the low and intermediate redshift bins for our sample (i.e., $0.4 \leq z \leq 1.7$), we measure an as-is median (i.e., not inclination-corrected) scale height of 0.91 $\pm$ 0.05 kpc---in good agreement with the low redshift average from \citet{2017ApJ...847...14E}. Moreover, the high redshift average is consistent with our measurements in the same redshift bin (0.96 $\pm$ 0.07 kpc). For both the sample studied in this paper and in \citet{2017ApJ...847...14E}, there is no significant evidence for a redshift evolution in the average scale height from $z\sim2.5$ to $z\sim0.4$.

\begin{table}

\begin{center}
\begin{tabular}{c|c|c|c}
\hline
$z$ & z$_{\mathrm{0, \mathrm{median}}}$ & z$^{\mathrm{incl-corr}}_{\mathrm{0, \mathrm{median}}}$ & $\Delta$z$_{\mathrm{0}}$ \\
&\multicolumn{3}{c}{(kpc)}\\
\hline\hline
$1.7 < z \leq 2.5$ & 0.96 $\pm$ 0.07 & 0.75 $\pm\,$0.05 & 0.47\\ 
(N = 198)&  & & \\ 
$0.8 < z \leq 1.7$ & 0.88 $\pm$ 0.04 & 0.69 $\pm\,$0.03 & 0.40\\ 
(245)&  & & \\ 
$0.4 \leq z \leq 0.8$ & 1.15 $\pm$ 0.15 &  0.90 $\pm\,$0.12 & 0.80\\
(48)&  & & \\ 
\hline
$0.4 \leq z \leq 2.5$& 0.94 $\pm$0.04 & 0.74 $\pm$ 0.03& 0.35\\
(491)& & & \\ 
\end{tabular}
\end{center}
\caption{The population median (z$_{\mathrm{0, \mathrm{median}}}$), inclination-corrected median (z$^{\mathrm{incl-corr}}_{\mathrm{0, \mathrm{median}}}$), and scatter ($\Delta$z$_{\mathrm{0}}$) of the galaxy scale height as measured at one effective radius is reported for different bins in redshift (top three rows) and for the full sample (bottom row). The scatter is defined as ($\mathrm{z}_{\mathrm{0, 84th}} - \mathrm{z}_{\mathrm{0, 16th}}$)/2. The reported uncertainties are the standard error on the median.}\label{table:z0_results}

\end{table}

In Figure \ref{fig:today}, we compare the distribution of the scale heights of our high-redshift galaxy sample with those of a population-matched sample of disk galaxies in the local universe as measured from the Sloan Digital Sky Survey (SDSS, see \S\ref{sec:SDSS}; \citealt{2014ApJ...787...24B}). The high-redshift distribution is shown in red, and the low-redshift distribution from SDSS is shown in blue. In addition, we show the range of estimates of the scale heights of the thin and thick disks of the Milky Way (\citealt{2016ARA&A..54..529B}, and references therein).

Figure \ref{fig:today} shows that the inclination-corrected {\emph{median}} of the high redshift population (0.74 $\pm$ 0.03 kpc) is consistent with the range of estimates (0.63--1.08 kpc) of the Milky Way thick disk. The thin disk of the Milky Way is significantly thinner than both the high redshift and the SDSS comparison samples. Importantly, we note that we are only able to measure the composite thicknesses of the external galaxies. While we {\emph{cannot}} rule out the presence of a thin disk in these galaxies, we {\emph{can}} conclude that the majority of their stellar light arises from a thick component.

While the median scale height of the high-redshift sample is consistent with that of the thick disk of the Milky Way, it is appreciably and statistically {\emph{smaller}} than that of the SDSS comparison sample (1.46 $\pm$ 0.04 kpc; \citealt{2014ApJ...787...24B}). 

We next consider differences in the shapes of the distributions. The low-redshift distribution not only includes an offset in the median, but is also wider than that of the high-redshift distribution. The (16, 50, 84)$^{\mathrm{th}}$ percentiles of the SDSS and high-redshift distributions are (0.87, 1.46, 2.25) kpc and (0.60, 0.94, 1.52) kpc, respectively. We compute the width of the two distributions (p$_{84}$ - p$_{16}$) as 1.39 and 0.91, respectively---the SDSS distribution is $\sim$53\% wider than that of our high redshift sample.

Together, the differences in the median {\emph{and}} width of the distributions imply that with cosmic time the disk population needs to both (i) on average become physically thicker and also (ii) develop higher variety with some galaxies remaining thin and some thickening. We also note that both distributions contain a tail towards higher scale heights, and that the tail extends further in the low redshift distribution. This implies that the thickness of the thickest galaxies at a given redshift increases with decreasing redshift. Interestingly, we do not detect such evolution in our sample (Figure \ref{fig:scales}). However, we note that \citet{2017ApJ...847...14E} report this exact result (a thickening of the thick-end of the population at a given redshift) for the observed I-band (rest NUV - blue light) of clumpy galaxies in the mass range of our sample ($\log\,M/\mathrm{M}_{\odot}\sim9-10$) over $z\sim3$ to $z\sim1$. This raises an important question: do the thickest galaxies (those that comprise the tail) at high redshift evolve into the thickest galaxies in the present-day? This is a question that will need to be answered with cosmological numerical simulations that both incorporate high star particle resolution (e.g., \citealt{2013ApJ...773...43B, 2019ApJ...873..129P, 2021MNRAS.503.1815B, 2021MNRAS.502.1433M}) and which can model large populations of galaxies.


As a whole, we draw two conclusions: (1) disks as thick as the Milky Way are established as early as cosmic noon at $z\sim2$, but also that (2) these high-redshift stellar disks are as thin or thinner than their (expected) descendant galaxies today. This suggests that the thickest components of today's galaxy disks start thick, and subsequently thicken at later times. This is consistent with the scenario outlined in \citet{2021MNRAS.503.1815B}, where stars are formed in thicker disks (with higher velocity dispersion) at higher redshifts {\emph{and also}} subsequently thicken at later times.

We do not detect an evolution in the scale height across our redshift range. However, we note that (as seen in the distribution of points in Figure \ref{fig:scales}) we lack the sampling at $z\,<\,1$ to make a strong statistical statement on late time evolution. We postulate that the population-wide thickening that is inferred in Figure \ref{fig:today} must occur in the period of cosmic time where our sample loses statistical power---after $z\,\sim\,1$ or in the last 8 Gyr.

\section{Conclusion} \label{sec:conclude}
Using {\emph{Hubble Space Telescope}}/ACS+WFC3 imaging of the GOODS-S galaxy field, we measure the rest-optical scale heights of \numsample{} galaxies spanning $0.4\,<\,z\,<\,2.5$ and $9\,<\,\log\,M_{*}/M_{\odot}\,<\,11$. We use these measurements to track the redshift evolution of the composite thicknesses of stellar disks to cosmic noon. We then compare our results with a population-matched sample of disk galaxies today from the Sloan Digital Sky Survey (SDSS; \citealt{2014ApJ...787...24B}). Our main conclusions are as follows:
\begin{itemize}

    \item  We measure a median intrinsic (inclination-corrected) scale height of 0.74 ($\pm0.03$) kpc and a population scatter of 0.35 kpc. This median scale height is consistent with the range of estimates of that of the Milky Way's thick disk today ($\sim$0.6 - 1.1 kpc). This indicates that disks that are as thick as the Milky Way's thick disk are in place at early cosmic times. 

    \item Comparing with the distribution of scale heights of local disks measured from SDSS \citep{2014ApJ...787...24B}, we find that the high-redshift population is {\emph{on average}} physically thinner than their population-matched descendent disks today. This indicates that the  disk population must on average thicken towards the present day. Moreover, the width of the low redshift distribution is larger than that of the high-redshift population (by 53$\%$), indicating that the galaxy population must develop a higher variety in thickness with cosmic time.

    \item From $z\,\sim\,2.5$ to $z\,\sim\,0.4$, we find no evidence for an evolution in the median and scatter of the scale heights of the galaxy population with redshift. We suggest that the bulk of the scale height evolution that is implied by the comparison with SDSS above must occur at cosmic times later than $z\sim1$, where our sample loses statistical power.
    
\end{itemize}

In brief, our results indicate that the stellar disks of galaxies {\emph{both}} start thick and subsequently thicken with time.

With the near-infrared imaging capabilities of {\emph{JWST}}/NIRCam + NIRISS and imaging observations from a number of large public programs available (e.g., CEERS, PRIMER, COSMOS-Web, UNCOVER), it is now possible to extend this study to earlier cosmic times at $z\,>\,3$. Early findings from these and other programs are indicating a surprisingly high fraction of {\emph{morphologically-}regular} stellar disks at these redshifts \citep{Robertson22, Kartaltepe22, 2022ApJ...938L...2F, 2022arXiv221001110F}. It is not yet clear how the detailed physical structures of these galaxies are connected with the structures of the galaxy populations studied in this paper at later cosmic times.

\begin{acknowledgments}
\section{Acknowledgements}
This work is supported through the Hubble Space Telescope program number AR-15052. Support for Program number AR-15052 was provided by NASA through a grant from the Space Telescope Science Institute, which is operated by the Association of Universities for Research in Astronomy. This work is based on observations taken by the 3D-HST Treasury Program (GO 12177 and 12328) with the NASA/ESA HST, which is operated by the Association of Universities for Research in Astronomy, Inc., under NASA contract NAS5-26555. This research made use of Astropy (http://www.astropy.org) a community-developed core Python package for Astronomy \citep{astropy:2013, astropy:2018}.

\end{acknowledgments}

%

\vspace{5mm}





\appendix
\section{Quantifying The Bias of Inclination on the Population Statistics} \label{sec:biases}
In this study, we measure the scale heights of a sample of galaxies that are selected to be ``edge-on". We define a galaxy to be edge-on if its {\emph{HST}}/F160W photometric axis ratio is $< 0.4$. In practice, every galaxy in the sample carries a finite inclination to our line-of-sight. Due to projection, this small inclination will artificially inflate the observed scale height. Ideally, we would correct each galaxy's scale height based on its known inclination. However, from the projected 2D images alone there is a degeneracy between disk thickness and galaxy inclination---we cannot distinguish between thick galaxies and inclined galaxies. However, we {\emph{are}} able to correct the population median scale height given the known population-averaged inclination. To do this, we rely on the known distribution of inclinations that a random sample of disks will take with respect to an arbitrary line-of-sight. In this appendix, we aim to quantify the bias of inclination on our population statistics by fitting a suite of {\emph{randomly-inclined}} disk models using the same procedures adopted in this paper (\S{\ref{sec:methods}}).

First, we generate three toy exponential disk galaxies in 3D ($\rho(r)\propto\,e^{-r/r_{d}}$). The three toy galaxies differ only in the intrinsic ratio ($q$) of their scale height to scale length ($r_d$), taking on $q=$ 0.15, 0.25, and 0.35, respectively. For each toy galaxy, we progressively incline and project it to a mock observer that is fixed in space. We sample an inclination array that follows a flat distribution in the cosine of the inclination angle. This is the distribution expected for randomly-inclined galaxies. In doing so, we create a suite of inclined projections. 

\begin{figure}
\begin{center}
\includegraphics[width=\columnwidth]{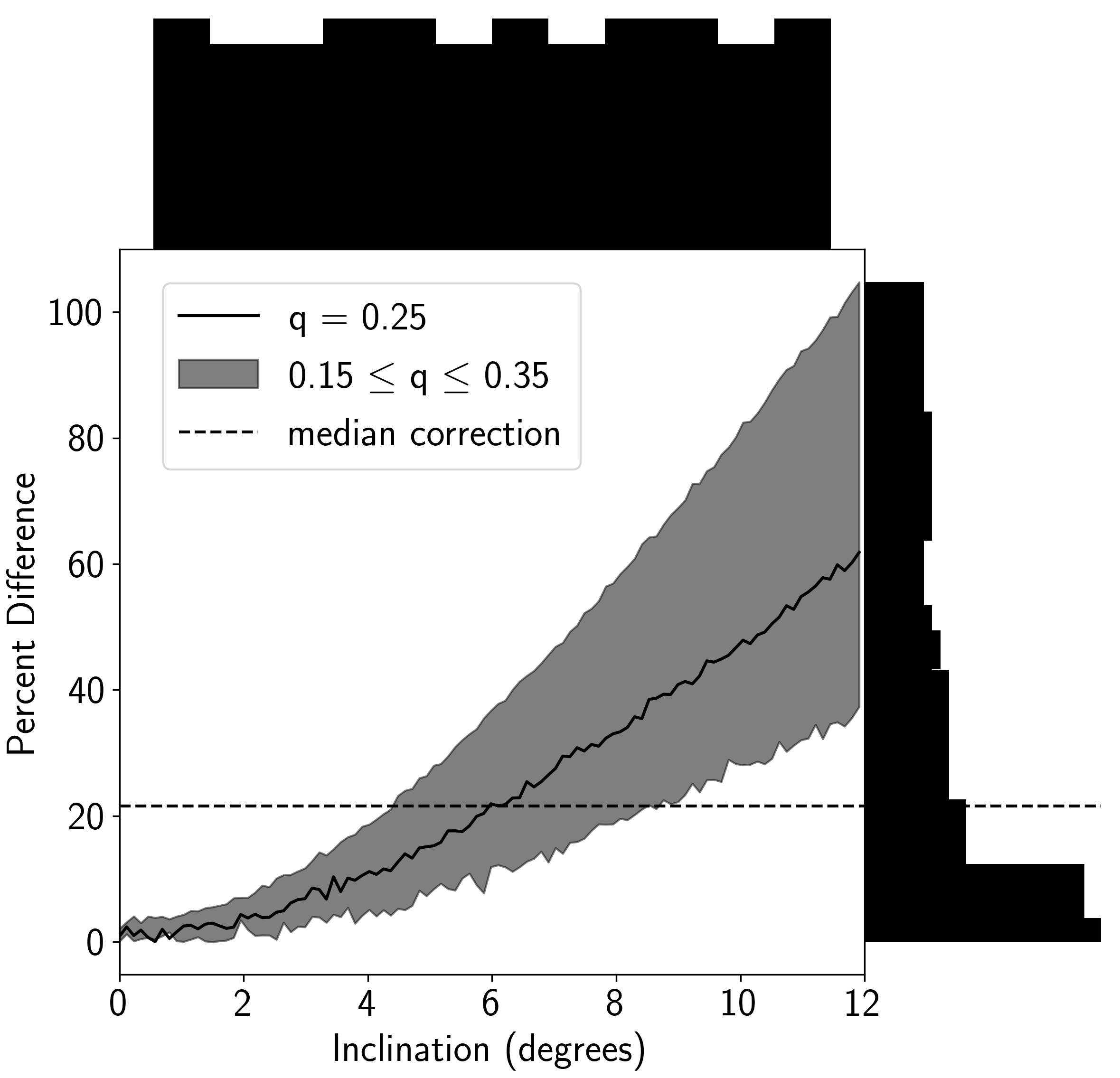}
\caption{This figure shows the percentage difference between the intrinsic and recovered scale heights for toy models of disk galaxies that are randomly inclined to our line-of-sight. The toy models have ratios ($q$) of intrinsic scale height to scale length that vary between 0.15 and 0.35 The horizontal line represents the median of the models, 22\%. We use the results shown here to correct the population median of the observed scale heights measured in this paper.}
\label{fig:incl}
\end{center}
\end{figure}

We then impose our selection criterion so that we only retain nearly edge-on galaxies as defined in this paper---those with a projected axis ratio of $<0.4$. For each galaxy that survives this cut, we perform the sech$^2$ fitting procedure described in \S \ref{sec:methods}. We take the scale height to be the median of the posterior of the model. By comparing the intrinsic scale height (which is fixed for all galaxies in our toy sample) with what we recover using the fits, we quantify the effect of inclination on the population statistics studied in this paper---notably the median. Figure \ref{fig:incl} shows how the percent difference between the intrinsic scale height of the toy galaxies and that recovered from our fitting procedure varies as a function of galaxy inclination and $q$. The median correction of the toy population is marked with a dashed line.

As the inclination of the toy disk model with respect to the observer increases, the percent difference between the intrinsic and recovered scale height also increases. Galaxies with inclinations larger than 12 degrees do not survive our selection criterion for the thickest galaxy models considered ($q = 0.35$). The solid line in Figure \ref{fig:incl} shows the $q = 0.25$ model. The shaded region shows the scatter, encompassing the lowest and highest axis ratios considered ($q = 0.15-0.35$). We calculate the median percentage difference between the intrinsic and recovered scale heights for all three axis ratios. The median of these medians is the dashed line, 22\%. We adopt this factor to correct the median of the observations studied in this paper. 

In conclusion, we derive a 22\% median difference between the intrinsic scale heights and those recovered for randomly-inclined disk galaxies that pass our selection criterion. This indicates that, on average, the scale heights for galaxies in our sample are artificially inflated by $\sim\,22\%$ and we correct the median results shown in Figures \ref{fig:scales} and \ref{fig:today} by this factor.

\begin{figure*}
\begin{center}
\includegraphics[width=\textwidth]{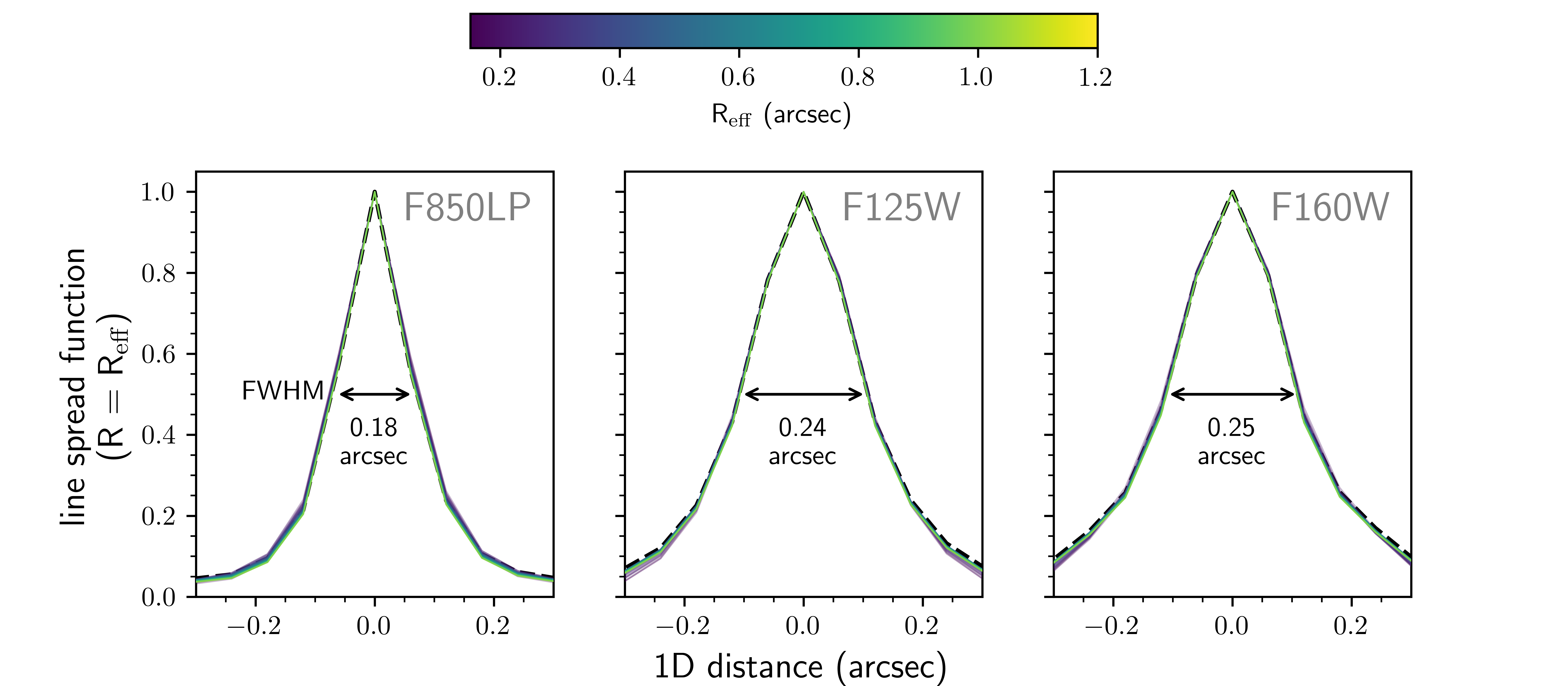}
\caption{The line spread functions (LSF) that are used to model the observed surface brightness profiles in this paper are shown by the black dashed lines. A LSF is constructed for each of the three {\emph{HST}} filters used in this paper---ACS/F850LP, WFC3/F125, and WFC3/F160W. They are determined by convolving a 2D point-spread function with an infinitesimally thick disk model of a given size (color-coding). The LSF are applicable at one effective radius---where the measurements in this study are carried out. A suite of LSF for disk models of varying sizes are shown with the color-coded lines. There is negligible variation in the LSF between the disk models. The FWHMs of the LSFs range from 0.$\arcsec$18 - 0.$\arcsec$25, depending on the filter.}
\label{fig:psf_1D}
\end{center}
\end{figure*}

\section{Constructing the 1D Convolution Kernel}\label{sec:kernel}
To fit the 1D vertical surface brightness profiles, we construct and convolve a 1D line spread function (LSF) with a $\mathrm{sech}^2$ model (see \S\ref{sec:methods}). The 1D LSF describes how the shape of the blurring from the 2D point-spread function varies with the distance above and below the galaxy midplane. To create the appropriate 1D kernel for our observations, we follow the approach of \citet{2017ApJ...847...14E} and simulate a convolution between the 2D point-spread function and a line of infinitesimal thickness.

We first create a suite of 2D images of mock disk galaxies that are perfectly edge-on and infinitesimally thick. The mock images adopt the pixel scale of the {\emph{HST}} images (0.\arcsec06 per pixel side). Each mock galaxy has a surface brightness profile that follows a 1D S\'ersic model (with $n=1$; and normalized to a total flux of 1) with a given scale length. We create a suite of mock galaxies with scale lengths that vary from 0.$\arcsec$2 - 1.$\arcsec$5---generally reflecting the distribution of the real galaxy sample. The major axis of each mock disk is set to lie along the horizontal of the image and the disks span a single {\emph{HST}} pixel (i.e., they are a line). For each mock disk, we simulate a convolved image in each of the three filters that we use in this paper ({\emph{HST}}/ACS+WFC3-F850LP, F125W, F160W), convolving with the 2D PSF provided by the 3D-HST survey \citep{2014ApJS..214...24S} for each of the filters.

For each image of each mock disk, we extract the surface brightness profile above and below the midplane of the mock disk. The shape of the resulting surface brightness profile corresponds to the appropriate 1D projection of the 2D PSF for a given pair of disk model and filter. Figure \ref{fig:psf_1D} shows the line spread functions derived for the suite of toy models in each of the three {\emph{HST}} filters.

In a given filter (i.e., a given 2D PSF), we find that the shape of the extracted 1D PSFs are relatively similar ($<10\%$ variation in the width) from column-to-column and for mock disks of different scale lengths. For all of the fits described in this paper, and for each of the three filters, we adopt the 1D PSF defined at one effective radius from the center of the mock disk with scale length 0.$\arcsec$4.


\bibliography{Paper.bib}{}
\bibliographystyle{aasjournal}



\end{document}